\documentclass[prl,preprint,amsmath,amssymb]{revtex4-1}

\usepackage{graphicx}
\usepackage{dcolumn}
\usepackage{bm}
\usepackage{multirow}
\usepackage{float}
\usepackage{wrapfig}
\setcitestyle{super}
\usepackage[normalem]{ulem}
\usepackage{color}
\usepackage{xr}
\usepackage{booktabs}
\usepackage{longtable}
\makeatletter
\providecommand \@ifxundefined [1]{%
 \@ifx{#1\undefined}
}%
\providecommand \@ifnum [1]{%
 \ifnum #1\expandafter \@firstoftwo
 \else \expandafter \@secondoftwo
 \fi
}%
\providecommand \@ifx [1]{%
 \ifx #1\expandafter \@firstoftwo
 \else \expandafter \@secondoftwo
 \fi
}%
\providecommand \href@noop [0]{\@secondoftwo}%
\providecommand \href [0]{\begingroup \@sanitize@url \@href}%
\providecommand \@href[1]{\@@startlink{#1}\@@href}%
\providecommand \@@href[1]{\endgroup#1\@@endlink}%
\providecommand \@sanitize@url [0]{\catcode `\\12\catcode `\$12\catcode
  `\&12\catcode `\#12\catcode `\^12\catcode `\_12\catcode `\%12\relax}%
\providecommand \@@startlink[1]{}%
\providecommand \@@endlink[0]{}%
\providecommand \url  [0]{\begingroup\@sanitize@url \@url }%
\providecommand \@url [1]{\endgroup\@href {#1}{\urlprefix }}%
\providecommand \urlprefix  [0]{URL }%
\providecommand \selectlanguage [0]{\@gobble}%
\providecommand \bibinfo  [0]{\@secondoftwo}%
\providecommand \bibfield  [0]{\@secondoftwo}%
\providecommand \BibitemShut  [1]{\csname bibitem#1\endcsname}%
\let\auto@bib@innerbib\@empty

\usepackage{ulem}
\usepackage{verbatim}
\usepackage{color}

\begin{document}
\preprint{}

\title{Four-Dimensional Imaging of Lattice Dynamics using Inelastic Scattering}

\author{Navdeep Rana$^{1}$}

\author{Aditya Prasad Roy$^{2}$}
\author{Dipanshu Bansal$^{2}$}
\email{dipanshu@iitb.ac.in}             
\author{Gopal Dixit$^{1}$}
\email{gdixit@phy.iitb.ac.in}

\affiliation{$^{1}$Department of Physics, Indian Institute of Technology Bombay,
           Powai, Mumbai 400076, India }
\affiliation{$^{2}$Department of Mechanical Engineering, Indian Institute of Technology Bombay, Powai, Mumbai 400076, India}

\date{\today}



\begin{abstract}
Time-resolved mapping of lattice dynamics in real- and momentum-space is essential to understand better
several ubiquitous phenomena such as heat transport, displacive phase transition, thermal conductivity, and many more. In this regard, time-resolved diffraction and microscopy methods are employed to image the induced lattice dynamics within a pump-probe configuration. In this work, we demonstrate that inelastic scattering methods, with the aid of theoretical simulation,  are competent to provide similar information as one could obtain from the time-resolved diffraction and imaging measurements. To illustrate the robustness of the proposed method, our simulated result of lattice dynamics in germanium is in excellent agreement with the time-resolved x-ray diffuse scattering measurement performed using x-ray free-electron laser. For a given inelastic scattering data in energy and momentum space, the proposed method is useful to image in-situ lattice dynamics under different environmental conditions of temperature, pressure, and magnetic field. Moreover, the technique will profoundly impact where time-resolved diffraction within the pump-probe setup is not feasible, for instance, in inelastic neutron scattering.
\end{abstract}

\maketitle

\section{Introduction} 
Inelastic scattering of matter allows us to probe quasi-particles (QPs) such as phonons (quantized lattice vibrations), magnons (quantized spin excitations), and polarons~\cite{squires1996introduction, Willis2009, schulke2007electron, egerton2011electron}. These QPs have finite energy and lifetime that carries information about the intra- and inter-QPs interactions and their coupling strength, necessary to understand the materials' response to external stimuli. Thermal conductivity~\cite{Niedziela_2019, lindsay2020thermal}, heat capacity~\cite{Dove1993, Fultz2010, Bansal_2016_2}, and phase transitions~\cite{Dove1993, Bansal_2018,bansal2020magnetically} are among many of the material properties that are often described by directly invoking various QPs. Optical, x-ray, neutron, and electron scattering methods -- for example, Raman scattering, inelastic x-ray scattering (IXS), inelastic neutron scattering (INS), and electron energy loss spectroscopy (EELS) -- have been routinely employed to measure these QPs~\cite{schulke2007electron, squires1996introduction, egerton2011electron}. Typically these measurements are performed in the momentum and energy domains (${\bf k}$-$\omega$), and lack information on temporal dynamics, i.e., ${\bf k}$-$t$ and ${\bf x}$-$t$ imaging -- time evolution of momentum or real space coordinates which ranges from femto- to several nano-seconds~\cite{gaffney2007imaging, sciaini2011femtosecond}.

To image ${\bf k}$-$t$ and ${\bf x}$-$t$ dynamics, a pump-probe setup having two ultrashort pulses, where the duration of the probe pulse must be shorter than the characteristic timescale of motion that is under probe, are required. 
Thanks to tremendous technological advancement, it has become possible to generate ultrashort x-ray and electron pulses,~\cite{hartmann2018attosecond, emma2, morimoto2018diffraction} and image the lattice dynamics 
in ${\bf k}$-$t$ and ${\bf x}$-$t$ domains. For example, the time-resolved x-ray and electron diffraction within a pump-probe configuration are used to image the lattice dynamics in the ${\bf k}$-$t$ domain~\cite{trigo2013fourier, clark2013ultrafast, elsaesser2014perspective, bredtmann2014x, fritz2007ultrafast, siwick2003atomic, wall2012atomistic, brown2019direct}, and the ultrafast electron microscopy has recently been demonstrated for the imaging in the ${\bf x}$-$t$ domain at an unprecedented spatiotemporal resolution~\cite{flannigan20124d, cremons2016femtosecond, fu2017imaging}. However, similar advances have not been taken for neutron sources to produce an ultra-short neutron pulse for imaging the lattice dynamics~\cite{pomerantz2014ultrashort}. 
At this juncture, it is not straightforward whether one can employ neutron sources within a pump-probe setup with sufficient atomic-scale spatiotemporal resolution to image lattice or spin dynamics ~\cite{pomerantz2014ultrashort}.

In this work, we theoretically demonstrate that methods based on inelastic scattering are suitable to extract similar information as one could get from the time-resolved imaging of lattice dynamics in ${\bf k}$-$t$ or ${\bf x}$-$t$ domains. Our approach is general and equally applicable to IXS, INS, and EELS. In general, all these inelastic scattering based methods probe dynamical structure factor $S(\mathbf{k}, \omega)$, apart from pre-factors, in experiments. The inelastic scattering methods provide ${\bf k}$-$\omega$ resolved measurement of QPs and comprise a powerful way to investigate the correlated motion of atoms and electrons~\cite{schulke2007electron, squires1996introduction, egerton2011electron}. We should mention that in ideal conditions, irrespective of whether measurements are in ${\bf k}$-$\omega$, ${\bf k}$-$t$, or ${\bf x}$-$t$ domains, they provide similar information after coordinate transformation(s). However, in practice, one measurement domain may have an advantage over the other. For example, under static environmental conditions of temperature, pressure, or magnetic field, the four-dimensional (4D) ${\bf k}$-$\omega$ mapping of QPs is preferred because of its superior energy and momentum resolutions ($\sim$0.1\,meV and $\sim$0.5\,nm$^{-1}$)~\cite{Willis2009, Ehlers, ARCS, HERIX3}, from which one can readily extract the QP energy, group velocity, and linewidth. On the other hand, the ${\bf k}$-$t$ domain is useful for tracking the temporal evolution of atomic motions upon photoexcitation-induced structural phase transitions~\cite{wall2018ultrafast} or the measurement of long-wavelength phonon lifetime (of the order of tens of picoseconds, which is not easily accessible in the ${\bf k}$-$\omega$ domain). Moreover, mapping the acoustic phonon wavefronts or the nucleation of waves from defects and interfaces in nanostructures is better suited for the ${\bf x}$-$t$ domain~\cite{cremons2016femtosecond, cremons2017defect}. 

In the following, we show that $S(\mathbf{k}, \omega)$ (obtained from experimental measurements or simulations)
 encodes all the essential information to image the lattice dynamics in the ${\bf k}$-$t$ and ${\bf x}$-$t$ domains after coordinate transformation without the causality violation. In particular, as we illustrate, our approach is well-suited to image the first-order states (i.e., emission or absorption of a single phonon at ${\bf q}\simeq 0$ from inelastically scattering photons or disorder-activated continuum~\cite{carles1982new}) and second-order `squeezed' states~\cite{trigo2013fourier} in the ${\bf k}$-$t$ domain (squeezed states are generated in the entire reciprocal lattice immediately after pumping the sample with a visible or near-infrared pump pulse)~\cite{trigo2013fourier, henighan2016control}. The temporal evolution and decay of the measured intensity from the change in phonon occupation at a given ${\bf k}$ point due to electron-electron, electron-phonon, and phonon-phonon scattering channels~\cite{trigo2013fourier,stern2018mapping,murphy2019evolution} are not explicitly included within the current framework. However, we consider the finite lifetime of first- or second-order states by including the phonon linewidths. Moreover, our methodology allows for ${\bf x}$-$t$ imaging of the coherent phonon dynamics from a point-like nucleation site or an extended defect. Our approach of imaging dynamics in the ${\bf x}$-$t$ domain can be directly compared with the electron microscopy data, as we demonstrate later by an example.

\section{Results and Discussion}

\subsection{Computational approach}

Silicon is used in the present work to demonstrate the proposed concept. The $S(\mathbf{k}, \omega)$ is simulated in the $({H}, {H}, {L})$ reciprocal plane following the same procedure as in our previous studies~\cite{FBao_2016_1,FBao_2016_2}. 
The range of energy transfer lies from  0 to 80\,meV with a step size  of 0.25\,meV (i.e., energy resolution), whereas the momentum transfer range varies from $(0,0,0)$ to $(4,4,7)$ reciprocal lattice units (r.l.u.)  with the step size of 0.025\,r.l.u. (see Supplementary Fig.~S3). Here, $a$ = 0.543\,nm is used as the lattice parameter of silicon. After calculating $S(\mathbf{k}, \omega)$ and  using fluctuation-dissipation theorem, the imaginary part of the response function $\chi(\mathbf{k}, \omega)$ is obtained as, $\textrm{Im}[\chi(\mathbf{k}, \omega)] = -\pi[S(\mathbf{k}, \omega) - S(\mathbf{k}, -\omega)]$. $S(\mathbf{k}, \omega)$ and $S(\mathbf{k}, -\omega)$ are related to each other by $S(\mathbf{k}, -\omega) = \textrm{exp}(-\beta \hbar \omega)~S(\mathbf{k}, \omega)$ with $\beta = (k_{\rm B}T)^{-1}$, where $k_{\rm B}$ is the Boltzmann's constant~\cite{schulke2007electron}. In the following all results are shown at $T$ = 300\,K. It is known that the real part of the response function is related to its imaginary part by the Kramers-Kronig relation,~\cite{jackson2007classical} 
\begin{equation}\label{eq1}
\textrm{Re}[\chi(\mathbf{k},\omega)] = \frac{1}{\pi} \mathcal{P} \int_{-\infty} ^{\infty} d \omega^{\prime}~ 
\frac{\textrm{Im}[\chi(\textbf{k},\omega ^{\prime})]}{(\omega^{\prime}-\omega)}.  
\end{equation}
Here, $\mathcal{P}$ represents the principal value of the integral. Fourier transform is performed to obtain $\chi(\mathbf{k}, t)$ from $\chi(\mathbf{k},\omega)$, which is complex in nature (see Supplementary Material for derivation). Equation~\eqref{eq1} ensures that $\chi(\mathbf{k}, t)$ = 0 for $t < 0$ and enforces the causality. It should be emphasized that direct Fourier transform of $S(\mathbf{k}, \omega)$ to $I(\mathbf{k}, t)$, i.e., $I(\mathbf{k}, t) = \hbar\int S(\mathbf{k}, \omega)\exp(\rm{i}\omega t)\,d\omega$,~\cite{squires1996introduction} is not the time evolution measured in time-resolved experiments, as $I(\mathbf{k}, t)$ violates causality. Note that, there is always a limitation in the energy resolution while measuring $S(\mathbf{k}, \omega)$ in an experiment. The discrete nature of the binned $\omega$ during experiment causes periodic nature of $\chi(\mathbf{k}, t)$. The time resolution of $\chi(\mathbf{k}, t)$ is estimated by Fourier law as $\Delta t = 2\pi\hbar/(80$\,meV) = 51.5\,fs and the time duration of the lattice dynamics is related to the range of the sampled values of $\omega$. Here, the time duration of the lattice dynamics in silicon is 8.2\,ps. We note that the $\Delta t$ and time duration are proportional to the energy transfer and energy binning (or resolution) and will vary from one experimental set-up to another. 
For example, Abbamonte and co-workers have imaged density disturbances in water  with attosecond time resolution ($\Delta t = 41.3$\,attoseconds) with incident x-ray energy and resolution of 100\,eV and 0.3\,eV, respectively~\cite{abbamonte2004imaging}.

\subsection{Response function in momentum domain}

\begin{figure}[h!]
\includegraphics[width= 14 cm]{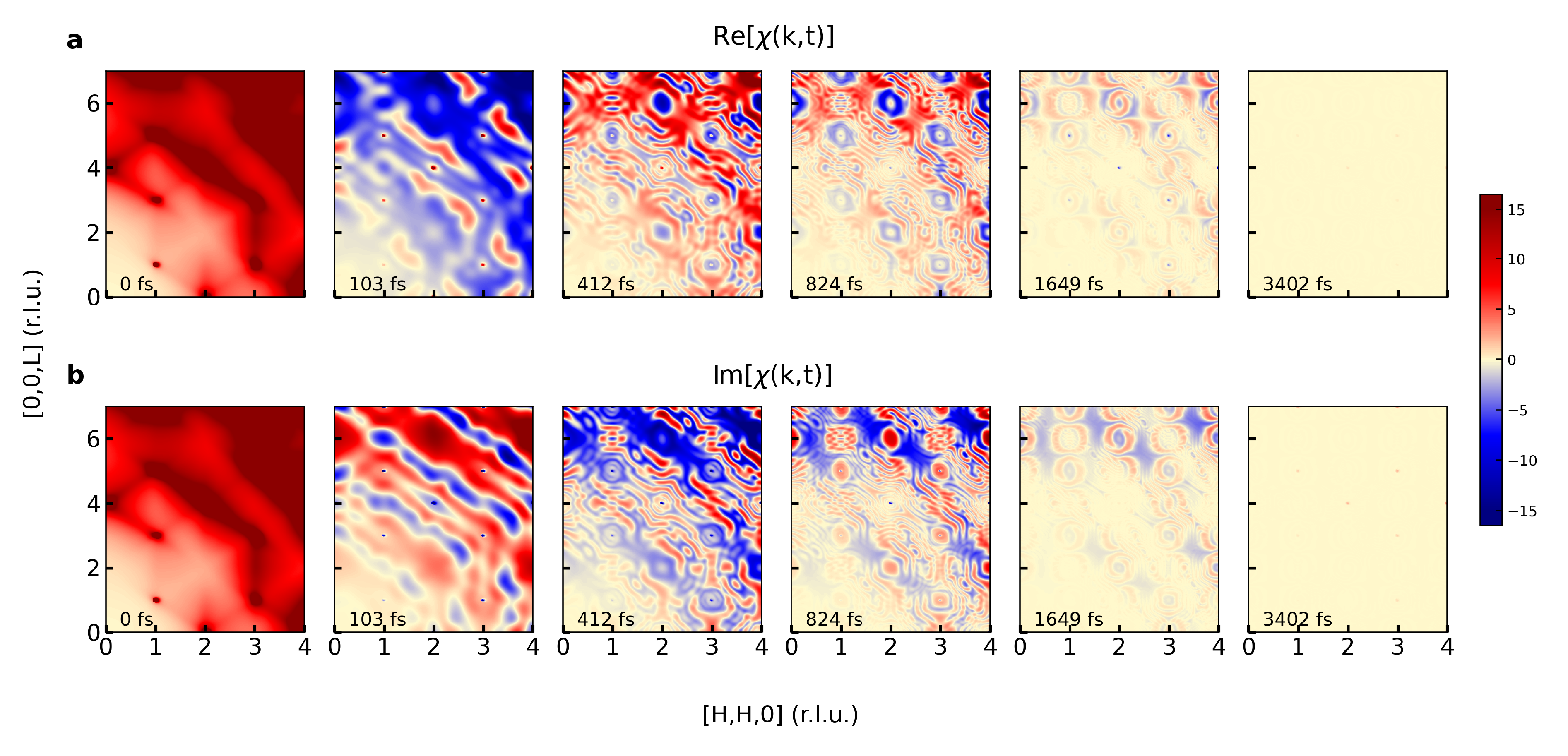}
\caption{{\bf{Response function $\chi(\mathbf{k}, t)$ at different instances.}} 
Snapshots of (a) real and  (b) imaginary parts of $\chi(\mathbf{k}, t)$ 
at 0, 103, 412, 824, 1649 and  3402 femtoseconds (fs). 
$\chi(\mathbf{k}, t)$ is shown in the $(H,H,L)$ reciprocal plane. 
The contour plots are normalized to their maximum intensity. 
The lattice dynamics dies out at longer time instances.}
\label{fig1}
\end{figure}

Figure~\ref{fig1} represents snapshots of the real and imaginary parts of the normalized $\chi(\mathbf{k}, t)$ in the upper  and lower panels, respectively. The dynamics can be seen as induced by the point-like source in ${\bf x}$,~\cite{abbamonte2010ultrafast} akin to the nucleation site in electron microscopy imaging. From this point-like source term, silicon absorbs the energy at $t = 0$, and phonon modes are generated in the entire reciprocal space (localized source in ${\bf x}$ is delocalized in ${\bf k}$). The snapshots are shown in the $(H,H,L)$ reciprocal plane for different time instances. We note that a (weak) optical pump pulse will also lead to the same $\chi(\mathbf{k}, t)$ snapshots, as the pump pulse will generate first- and second-order states in the entire reciprocal space. The first-order disorder-activated continuum is generated in the absence of perfect crystalline order~\cite{carles1982new}. In contrast, the second-order squeezed states are generated by the coupling of a photon (momentum ${\bf q}\simeq0$) with the two phonon modes of near-equal and opposite momenta (i.e., at $\mathbf{k}$ and $\mathbf{-k}$ due to the conservation of momentum)~\cite{henighan2016control}. Thus in the present study, the simulated snapshots can be considered to arise from either a point-like source in ${\bf x}$ or generation of first- and second-order states. As evident from Fig.~\ref{fig1}, these generated phonons propagate through the system and decay at different times for different $\mathbf{k}$ points according to their lifetime. It is possible to extract the lifetime of these phonon modes at different $\mathbf{k}$ values from the real and imaginary parts of the normalized $\chi(\mathbf{k}, t)$ as we discuss below.

\begin{figure}[h!]
\includegraphics[width= 14 cm]{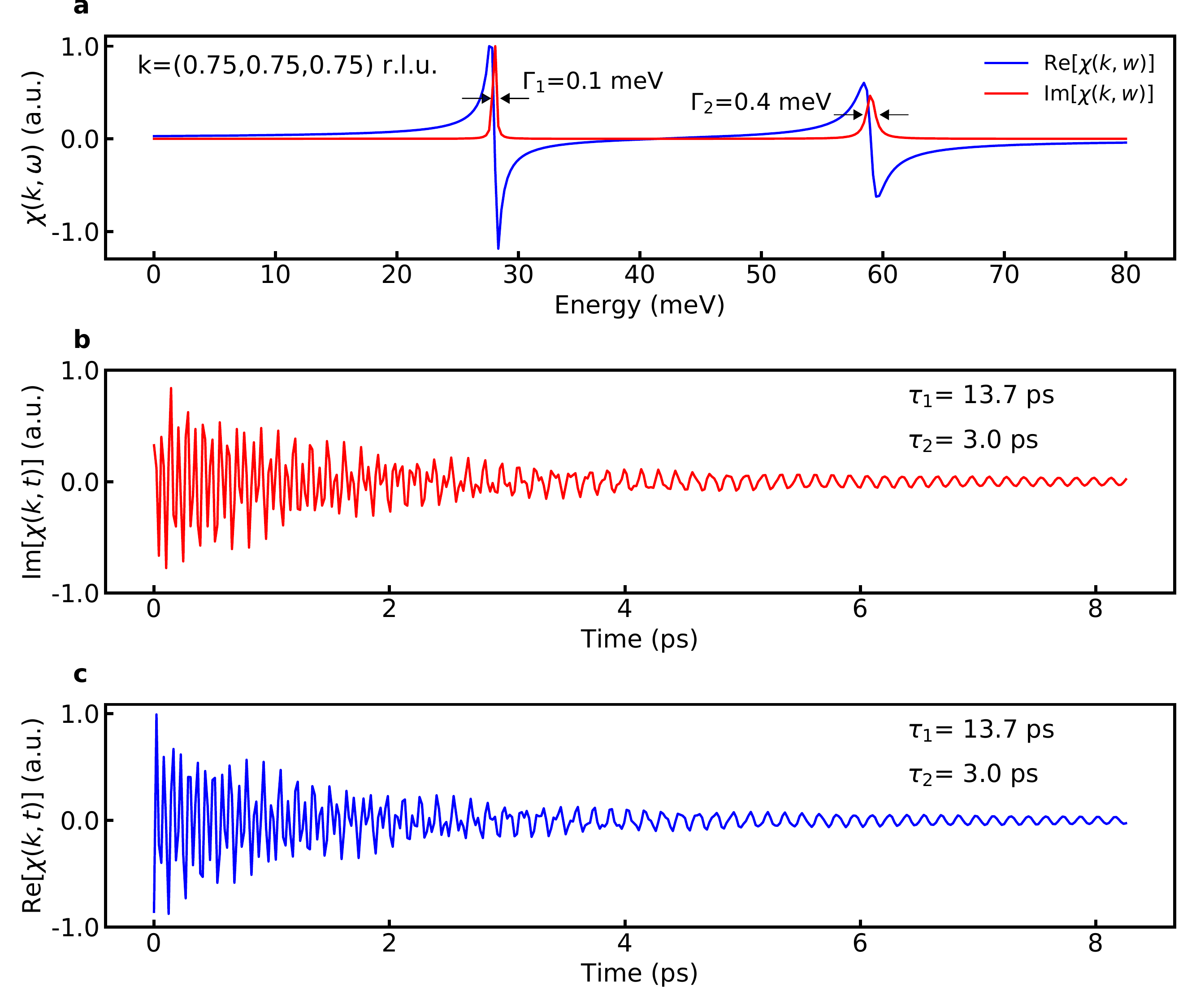}
\caption{{\bf Decay width and lifetime of phonon modes at particular $\mathbf{k}$ point.} 
(a) Real and imaginary parts of $\chi(\textbf{k}, \omega)$ at particular $\mathbf{k} = (0.75, 0.75, 0.75)$\,r.l.u. The decay widths corresponding to two active modes are $\Gamma_{1} = 0.1$\,meV and $\Gamma_{2} = 0.4$\,meV. (b) Imaginary part of $\chi(\textbf{k},t)$, which provides the lifetimes of the active phonon modes as $\tau_{1} = 13.7$\,ps  and $\tau_{2} = 3.0$\,ps. (c) Real part of $\chi(\textbf{k},t)$, which gives the same values of the lifetimes of the active phonon modes. All the quantities plotted in sub-plots are normalized.} 
\label{fig2}
\end{figure}

The real and imaginary parts of $\chi(\textbf{k}, \omega)$ at a particular $\mathbf{k} = (0.75, 0.75, 0.75)$ r.l.u.~is presented in Fig.~\ref{fig2}a. Because of the phonon polarization factor~\cite{squires1996introduction}, at this particular $\mathbf{k}$ value, two phonon modes have finite intensity at energy values $\omega_{1}$= 28 meV (longitudinal acoustic) 
and $\omega_{2}$=60 meV (longitudinal optic). 
The full width at half maximum (FWHM) at these two energy values are $\Gamma_{1} = 0.1$ meV and $\Gamma_{2} = 0.4$ meV (obtained from first-principles simulation of Silicon~\cite{carrete2017almabte}), which are known as decay widths of these modes. To extract the lifetime of these modes, $\chi(\textbf{k}, t)$ is calculated from $\chi(\textbf{k}, \omega)$. The imaginary and real parts of $\chi(\textbf{k}, t)$  are shown in Figs.~\ref{fig2}b and c, respectively. The lifetime ($\tau$) of both the active modes is obtained by fitting the exponentially decaying sinusoidal oscillation at the given $\mathbf{k}$ value, and are $\tau_{1} = 13.7$ and $\tau_{2} = 3.0$\,ps. The fitted $\tau$ values are consistent with the values expected form its inverse relationship with $\Gamma$, i.e., $\Gamma = 1/\pi\tau$, confirming the accuracy of our implementation. Also, the interferences visible in Figs.~\ref{fig2}b and c are due to the presence of two active modes. 

\begin{figure}[h!]
\includegraphics[width= 18 cm]{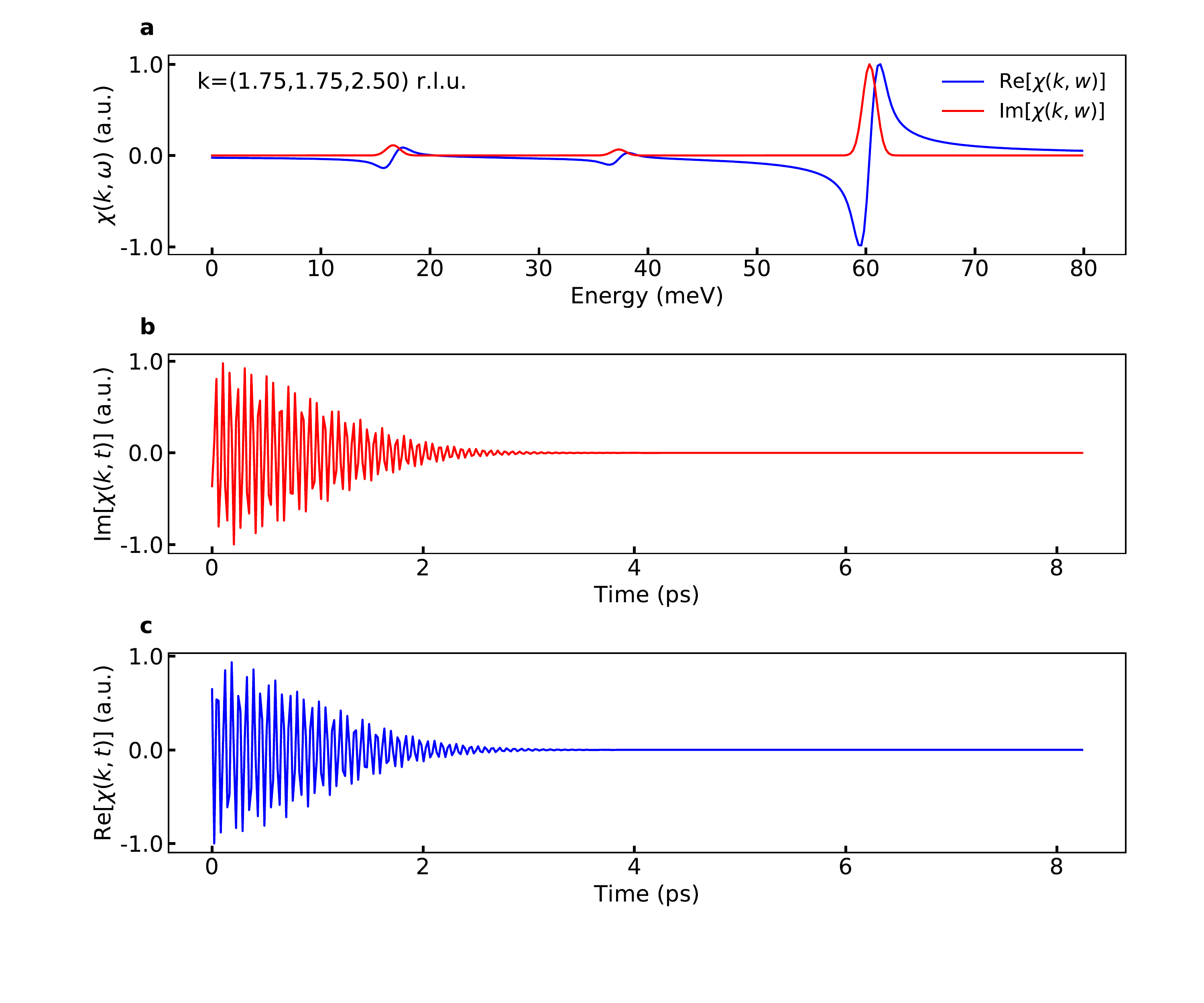}
\caption{{\bf Decay width and dynamics of phonon modes excited by an extended source in ${\bf x}$.} 
(a) Real and imaginary parts of $\chi(\textbf{k}, \omega)$. (b) Imaginary and (c) real parts of $\chi(\textbf{k},t)$. Here, the phonon modes at a specific $\mathbf{k}$ value of $\mathbf{k_0} = (1.75,1.75, 2.50)$\,r.l.u.~are excited by an extended source in ${\bf x}$. All the quantities plotted in sub-plots are normalized.}
\label{fig3}
\end{figure}

Till now, we have discussed dynamics induced by the point-like disturbance in ${\bf x}$, or first- and second-order states that lead to the excitation of all phonon modes in the reciprocal space. Let us analyze how $\chi(\textbf{k}, t)$ manifests when the extended source in ${\bf x}$ induces the dynamics, which excites the phonon modes active at a single $\mathbf{k}$ value. A general time-dependent extended external source $n_{ext}({\bf x},t)$ can be treated as a Gaussian envelop in the ${\bf k}$-$\omega$ domain with standard deviation $\sigma$ controlling the spatial extent of the source. The induced dynamics $n_{ind}$ can subsequently be modeled as 
\begin{equation}\label{eq2}
n_{ind}(\mathbf{k}, \omega) = \frac{4 \pi}{k^{2}} 
\frac{1}{ \sqrt{2 \pi \sigma^{2}}} ~\textrm{exp}{[{-(\mathbf{k}-\mathbf{k_0})^{2}}/{2 \sigma^{2}}]} 
~\chi(\mathbf{k}, \omega),
\end{equation}
where $\mathbf{k_0}$ is the mean $\mathbf{k}$-value at which phonons are excited. The real and imaginary parts of $\chi(\textbf{k}, \omega)$ at $\mathbf{k_0} = (1.75, 1.75, 2.50)$\,r.l.u.~showing finite intensity for the three phonon modes are shown in Fig.~\ref{fig3}a. Since momentum transfer ${\bf k}$ and phonon wavevector ${\bf q}$ at $\mathbf{k_0}$ are not entirely parallel or perpendicular to each other, the three modes have mixed transverse and longitudinal character. 
We emphasize that for an infinitely extended source as considered here, dynamics at all other ${\bf k}$ points are zero. As one can observe from Figs.~\ref{fig3}b and c, finite line-widths of the phonon modes allow them to decay in few picoseconds. Note that we do not explicitly include the phonon scattering channels in our approach. When included, these scattering channels will increase the phonon population with time at other ${\bf k}$ points that satisfy the momentum and energy conservation~\cite{trigo2013fourier,stern2018mapping,murphy2019evolution}. 

\subsection{Response function and time-resolved diffuse x-ray scattering}
To know how well-grounded the discussed method of extracting $\chi(\mathbf{k}, t)$ from $S(\mathbf{k}, \omega)$ (obtained from inelastic scattering measurements or simulations) is, we will compare our simulated results with the measured $\chi(\mathbf{k}, t)$. Trigo \emph{et al.} have performed time-resolved diffuse x-ray scattering on germanium~\cite{trigo2013fourier}. In that experiment, an optical pump pulse of 800\,nm with a nominal width of 50\,fs was used to generate correlated pairs of phonons with equal and opposite momenta at ${\bf k}$ and ${\bf -k}$, i.e., the squeezed states having $\langle { u}(t)\rangle = 0$. Here $u$ is the atomic displacement, and $t$ denotes the pump-probe delay time. A 50\,fs x-ray pulse having 10\,keV photon energy from Linac Coherent Light Source (LCLS) was used to probe diffuse scattering from the squeezed states at various pump-probe delay times. The temporal evolution of the equal-time correlation function was probed~\cite{trigo2013fourier}. Due to the time-resolved nature of the experiment, the time evolution of the squeezed states and the anharmonic decay of phonons is seen in the ${\bf k}$-$t$ domain. Note that for the squeezed states, the diffuse scattering intensity oscillates at twice the phonon frequency~\cite{trigo2013fourier}.

\begin{figure}[h!]
\includegraphics[width= 14 cm]{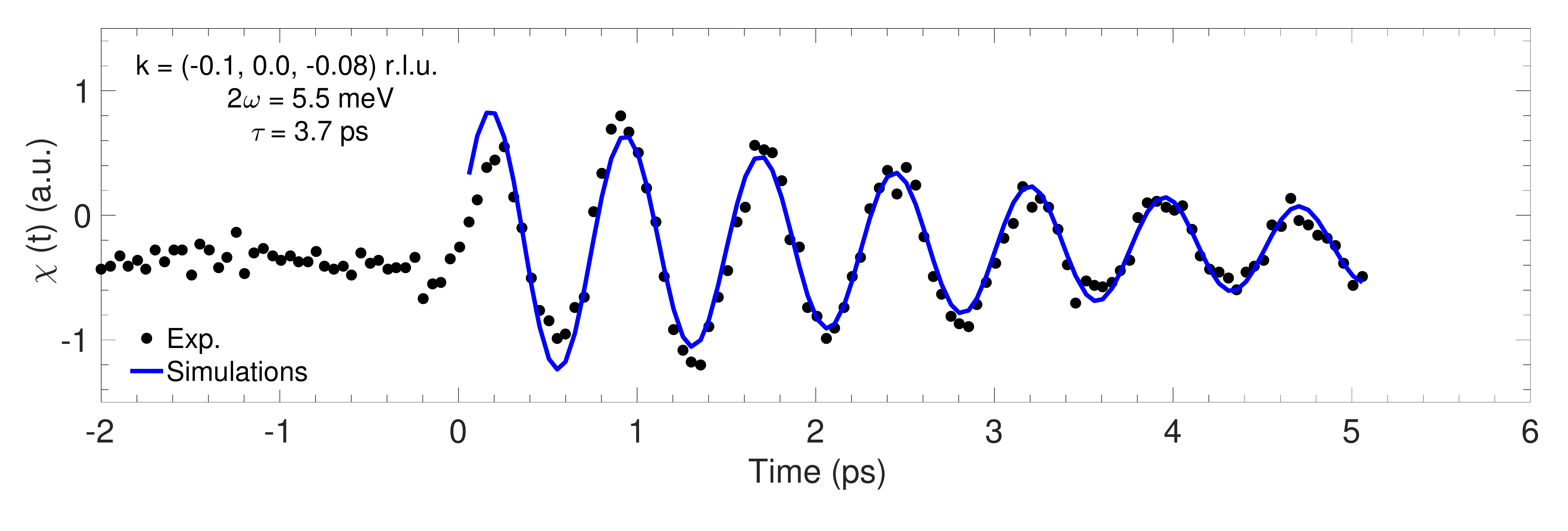}
\caption{ { \bf Comparison  of experimental data with simulated $\chi(t)$ for germanium.}  Normalized difference intensity of the diffuse scattering at $\mathbf{k} = (-0.10, 0.00, -0.08)$\,r.l.u.~is from the experiment presented in Ref.~\citenum{trigo2013fourier} (black color), and our simulated $\chi(t)$ is shown by blue color.} 
\label{fig4}
\end{figure}

Figure~\ref{fig4} presents the comparison of the experimental data from Ref.~\citenum{trigo2013fourier} with our simulated result of $\chi(\mathbf{k}, t)$ for germanium. To demonstrate the merit of our work, we have chosen the data at $\mathbf{k} = (-0.10, 0.00, -0.08)$\,r.l.u.~only. Since $\chi(\mathbf{k}, t)$ is dependent on the phonon mode polarization, the calculated intensity at $\mathbf{k}$ is due to the oscillations at $2\omega\sim$ 5.5\,meV. In the experiment, the normalized difference intensity was shown, which provides the time-resolved value of the equal-time correlation function at $\mathbf{k}$ and $\mathbf{-k}$~\cite{trigo2013fourier}. As reflected from the figure, the present simulated result is an excellent agreement with the experimental data for $\tau\sim$ 3.7\,ps. Such a large value of $\tau$ for dispersive phonon modes, which is straightforward to obtain from ${\bf k}$-$t$ domain measurements, is not so easy to extract from INS or IXS measurements owing to the finite instrument resolution in ${\bf k}$ and $\omega$~\cite{Ehlers, ARCS, HERIX3}.

In spite of the excellent agreement, an important question arises: how the two different methods -- time-resolved diffuse x-ray scattering and inelastic x-ray scattering, yield the same information. In Ref.~\citenum{trigo2013fourier}, it is mentioned that time-resolved diffuse x-ray scattering probes equal-time density-density correlation function: $\langle \hat{n}(\mathbf{-k},t)~\hat{n}(\mathbf{k},t) \rangle$ with $\hat{n}$ as a density operator~\cite{dixit2014theory, dixit2012imaging}. On the other hand, it is well-established that inelastic x-ray scattering probes density-density correlation function at different time: $\langle \hat{n}(\mathbf{-k}, t)~\hat{n}(\mathbf{k}, 0) \rangle$, which is related to the Van Hove correlation function~\cite{schulke2007electron, abbamonte2010ultrafast, van1954correlations}. However, there is no contradiction as the experiment was performed without energy resolution and the presented data were energy integrated~\cite{trigo2013fourier}. The density-density correlation function at different time, probed by inelastic x-ray scattering, reduces to equal-time density-density correlation function in the case where energy resolution is lacking. Therefore, without energy resolution, time-resolved diffuse x-ray scattering and inelastic scattering yield identical information: $\langle \hat{n}(\mathbf{-k},t)~\hat{n}(\mathbf{k},t) \rangle$, apart from a pre-factor (see Supplementary Material for derivation).

\subsection{Response function in real-space}

Not only time-resolved scattering methods in a pump-probe configuration provide the temporal evolution of correlation function, but also help us to visualize atomic motion (lattice dynamics) in the ${\bf x}$-$t$ domain~\cite{gaffney2007imaging, sciaini2011femtosecond}, for example, as in the time-resolved electron microscopy experiments~\cite{flannigan20124d, cremons2016femtosecond, cremons2017defect}. In the following, we demonstrate that momentum and energy-resolved inelastic scattering signal also provides the snapshots of the lattice dynamics in the ${\bf x}$-$t$ domain. For this purpose, we need to perform one more Fourier transform from momentum space to real space to obtain $\chi(\textbf{x}, t)$ from $\chi(\textbf{k}, t)$. Following Fourier relation, the spatial resolution along the $[H,H,0]$ direction is estimated as $\Delta \mathbf{x} = {2 \pi}/({6.5}$\,\AA$^{-1})$ = 0.96\,\AA\,, whereas along the $[0,0,L]$ direction is $\Delta \mathbf{x} = {2 \pi}/({8.1}$\,\AA$^{-1})$ = 0.78\,\AA. The spatial extent of the dynamics ranges from $(0,0)$ to $(153,108)$\,\AA. 
Similar to the time resolution and duration, the spatial resolution and extent are governed by momentum transfer and momentum resolution of the measurements/simulations.

\begin{figure}[h!]
\includegraphics[scale=0.8]{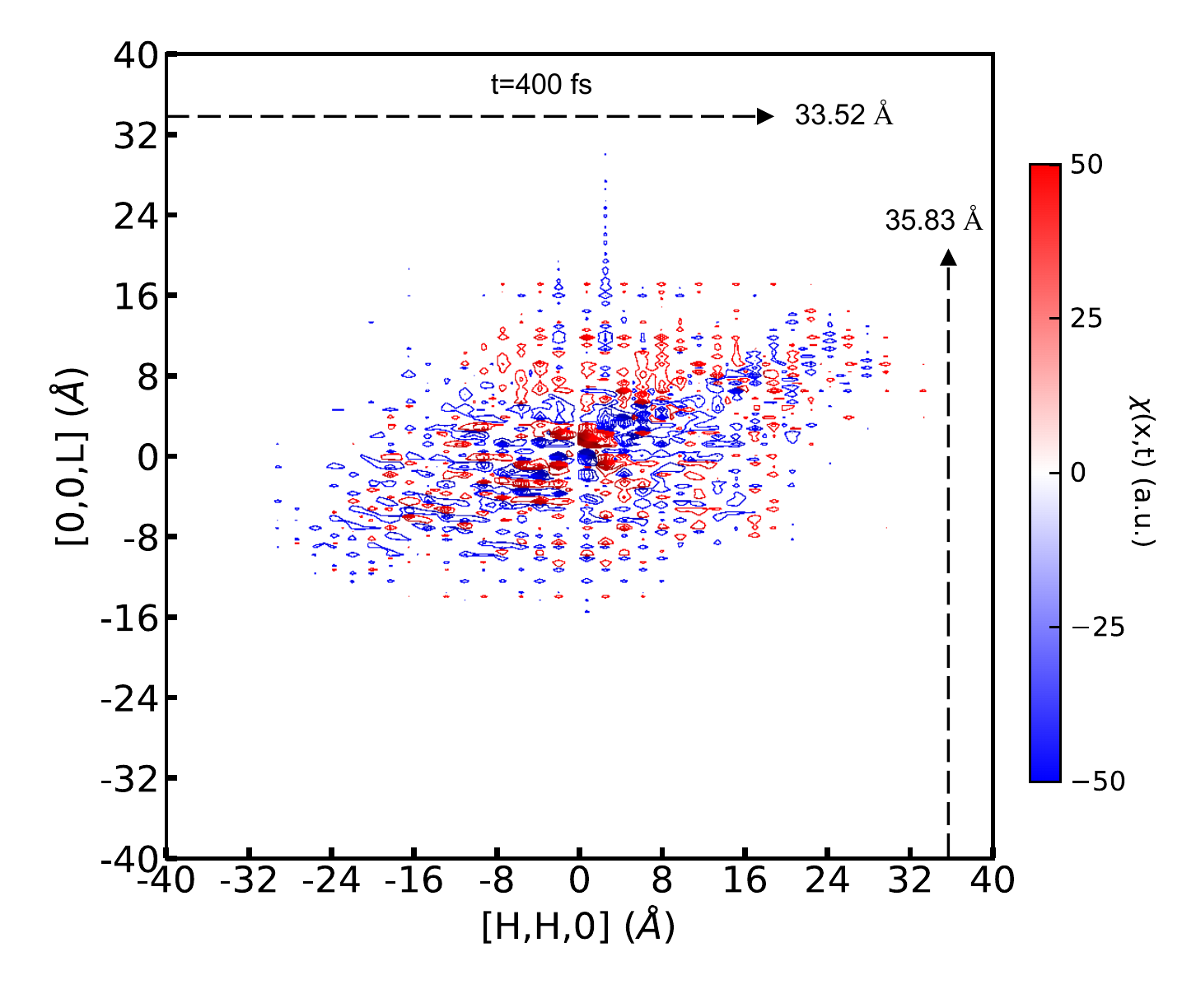}
\caption{{\bf Snapshot of 
$\chi(\mathbf{x}, t)$ in the $(H,H,L)$ plane at $t$ = 400\,fs for silicon.} 
The dotted lines represent the extent to which there are disturbances in $[1,1,0]$ and $[0,0,1]$ directions, which is given by the dispersion of longitudinal acoustic mode at the zone-center.}
\label{fig5}
\end{figure}

The snapshot of $\chi(\textbf{x}, t)$ indicates how and where the disturbance, imparted at $t = 0$, has travelled in real space. $\chi(\textbf{x}, t)$ at $t$ = 400\,fs in the $(H,H,L)$ plane is shown in Fig.~\ref{fig5}. 
In this case, the dynamics is induced by a point source at ${\bf x} = (0,0,0)$ in silicon. 
As silicon crystal is anisotropic; therefore the phonon propagation is not spherical as generally assumed for the scattering of light waves from a point-like source. Instead, the anisotropic phonon group velocity governs the 3D extent of energy re-distribution after time $t$.
The extent of disturbance along a particular direction at any given time can be estimated from the maximum velocity given by the dispersion of longitudinal acoustic mode at the zone-center. In the present case, the maximum velocity along the $[H,H,0]$ direction is 8958\,ms$^{-1}$, whereas it is 8381\,ms$^{-1}$ along the $[0,0,L]$ direction.

\begin{figure}[h!]
\includegraphics[scale=0.75]{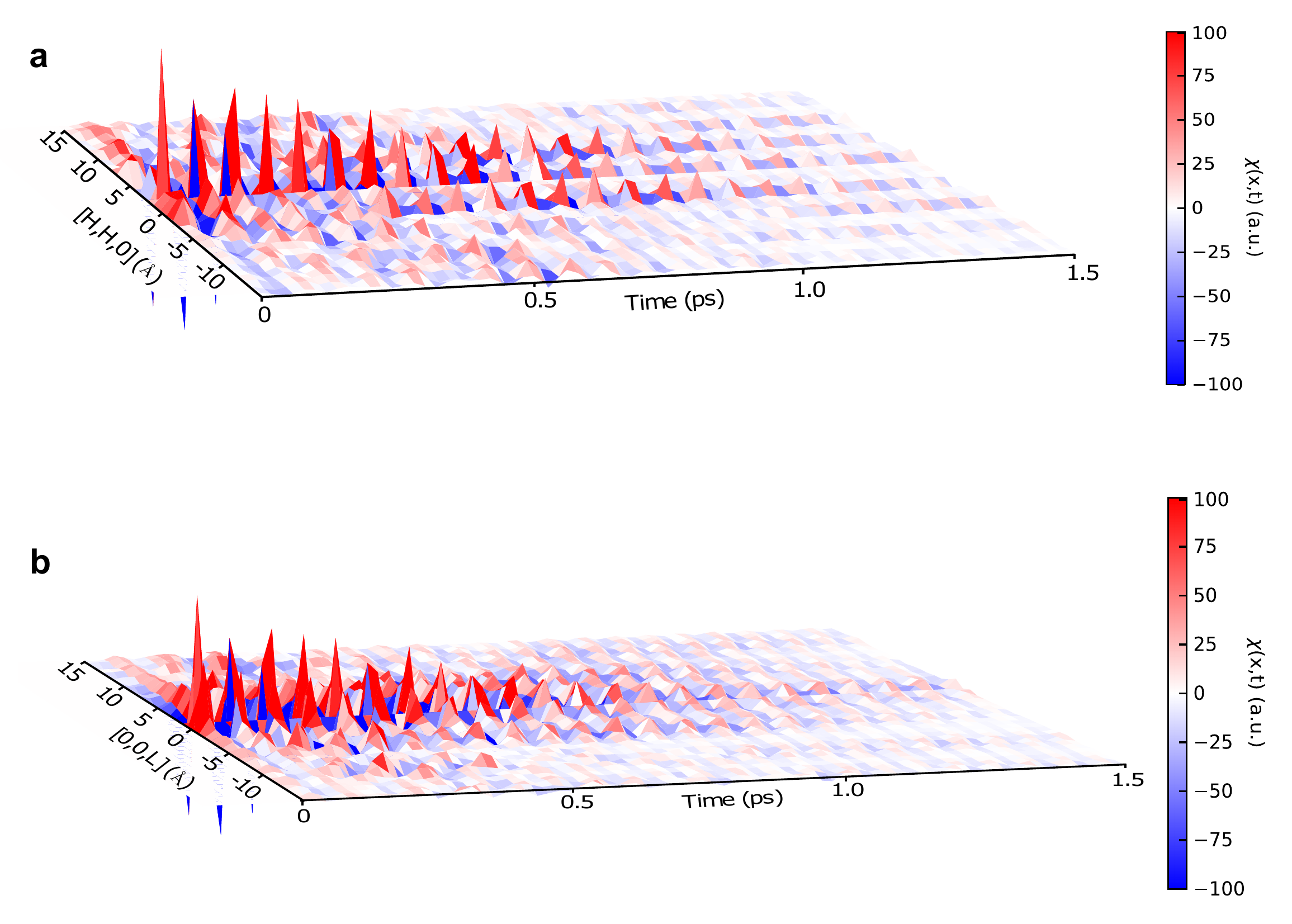}
\caption{{\bf Visualization of the lattice dynamics induced by a point source in silicon in the $({\bf{x},t})$ domain.} The dynamics 
(a) along the $[1,1,0]$ direction and (b) along the $[0,0,1]$ direction. The disturbance is still in the system at large instances, but the ripples' height is low. There is no dissipation of energy from the system.}
\label{fig6}
\end{figure}

Figure~\ref{fig6} presents the entire dynamics along $[1,1,0]$ and $[0,0,1]$ directions in upper and lower panels, respectively. Before the disturbance at $ t< 0$, the system is in equilibrium. At $t=0$, the system is struck with a negative disturbance. As a result of this disturbance, a positive recoil at the origin is generated, surrounded by a minimal positive build-up. As visible from the figure, the density-induced disturbance is propagating through the entire system as time evolves. At large time instances, the disturbance is still into the system, but the order is minimal due to the spread of energy into the system (no dissipation of energy from the system). Such time-resolved images can be captured nowadays using real-space femtosecond electron imaging, as recently demonstrated for the phonon nucleation and launch at a crystal step-edge in the WSe$_2$ flake~\cite{cremons2016femtosecond}. At this point, it is important to mention that Abbamonte and co-workers have employed a similar reconstruction method of $\chi(\mathbf{x}, t)$ from inelastic x-ray scattering data to visualize electron dynamics in various systems~\cite{abbamonte2004imaging, abbamonte2008dynamical, abbamonte2009implicit}.

\subsection{Practical challenges}

Despite the applicability of our approach to inelastic scattering methods, we should keep experimental and data analysis limitations in mind. For example, 4D momentum-, crystallographic direction-, and energy-resolved datasets of lattice dynamics are presently feasible with high-resolution ($\sim$1\,meV, or better) INS and IXS~\cite{Ehlers, ARCS, HERIX3}. In EELS, the energy resolution may always not be suitable for low energy phonons ($<$\, 30\,meV), as intense zero-loss peak masks the spectrum~\cite{krivanek2014vibrational, venkatraman2019vibrational}. Even in INS and IXS measurements, the scattering intensity varies among elements. Since x-ray scattering cross-section increases with the atomic number $Z$, phonon intensity of low $Z$ elements is considerably weak. Moreover, acquiring a complete 4D dataset using IXS with an array of monochromators will require a long measurement time (several days, as measurements at a few momentum transfers generally take 15 to 120 minutes). Similarly, in INS, several elements have small coherent scattering cross-section (i.e., H, V, Co) or high absorption coefficient (i.e., B, Cd); thus, phonon dispersion measurements remain challenging in materials with these elements. INS measurements also have kinematic constraints (all momentum and energy transfer are not accessible), and energy resolution, instead of being a constant number, depends on the phonon energy~\cite{squires1996introduction, Willis2009}. Deconvolution of energy resolution from the measured 4D dataset to extract the intrinsic details, for example, $\tau$, is not always straightforward~\cite{lin2016mcvine}. Recent advances in experimental techniques and data analysis are overcoming many of the limitations. Together with simulations, they can provide full or complementary information, as demonstrated in this study.

In summary, we have established that inelastic scattering methods have potential to image lattice dynamics with atomic-scale spatiotemporal resolution. Few tens of femtoseconds temporal and $\sim 1$\,\AA~spatial resolutions can be achieved during the reconstruction of lattice dynamics by utilizing the superior energy and momentum resolutions of inelastic scattering measurements. Our proposed theoretical method allows for direct imaging of lattice dynamics and enables us to extract the lifetime of selective phonon modes. Moreover, there is a flexibility to decide whether a single phonon mode or several phonon modes participate in the dynamics by choosing the disturbance source's spatial extent in the real-space. The excellent agreement between the present simulated result and the measured lattice dynamics in germanium provides confidence and robustness of the proposed method. We believe that the current approach will be an alternative to time-resolved diffraction methods to image lattice dynamics and beneficial to the situations where time-resolved diffraction is not easy to perform, such as neutron scattering and in-situ measurement conditions. 

\section{Methods}

\subsection{Dynamical structure factor calculation}
The dynamical structure factor $S(\mathbf{k}, \omega)$ was calculated using the following expression:
\begin{footnotesize}
\begin{align} \label{eq:DDCS1}
S(\mathbf{k}, \omega) \propto & \sum_{s}\sum_{\mathbf{\tau}} \frac{1}{\omega_{s}}  \left| \sum_{d} \frac{{f_d({\bf k})}}{\sqrt{M_d}} \rm{exp}(-W_d)\rm{exp}(i\mathbf{k}\cdot\mathbf{d})(\mathbf{k}\cdot\mathbf{e}_{ds})\right|^2\times \langle n_s  + \frac{1}{2} \pm \frac{1}{2}\rangle \delta(\omega \mp \omega_s)\delta(\mathbf{k}-\mathbf{q} -\mathbf{\tau}),
\end{align}
\end{footnotesize}
where $f_d({\bf k})$ is the form factor for atom $d$ (can be replaced by the neutron scattering length $\overline{b_d}$ for inelastic neutron scattering), $\textbf{k = k' - k''}$ is the wavevector or momentum transfer, and $\textbf{k''}$ and $\textbf{k'}$ are the final and incident wavevector of the scattered particle, respectively; $\textbf{q}$ is the phonon wavevector, $\omega_s$ is the eigenvalue and 
$\mathbf{e}_{ds}$ is the eigenvector of the phonon corresponding to the branch index $s$, $d$ is the atom index in the unit cell, $\tau$ is the reciprocal lattice vector, $\exp(-2W_d)$ is corresponding to the Debye-Waller factor, and $n_s = \left[\exp\left(\frac{\hbar\omega_s}{k_{\rm B}T}\right)-1\right]^{-1}$ is the Bose-Einstein occupation factor. The $+$ and $-$ sign in Eq.~\eqref{eq:DDCS1} correspond to phonon creation and phonon annihilation, respectively. The phonon eigenvalues and eigenvectors (simulation details are same as presented in Refs.~\citenum{FBao_2016_1} and~\citenum{FBao_2016_2}) in Eq.~\eqref{eq:DDCS1} were obtained by solving dynamical matrix using Phonopy~\cite{Phonopy_2015}. Phonon dispersion curves, $S(\mathbf{k}, \omega)$ and constant energy slices of $S(\mathbf{k}, \omega)$ of silicon and germanium are given in Supplementary Information. 

\section*{Data Availability}
Data that support the plots within this paper and other findings of this study are available from the corresponding authors upon reasonable request.

\section*{Code Availability}
Code that support the findings of this study are available from the corresponding authors on reasonable request.

\section*{Acknowledgements}
G.D. acknowledges fruitful discussion with Sucharita Giri. A.P.R. acknowledges the financial support from IRCC-IITB. D.B. thanks the financial support from MHRD-STARS under project no.: STARS/APR2019/PS/345/FS, and BRNS -- DAE under project no.: 58/14/30/2019-BRNS/11117. G. D. acknowledges support from Science and Engineering Research Board (SERB) India 
(Project No. ECR/2017/001460) and the Ramanujan fellowship (SB/S2/ RJN-152/2015).  

\section*{Author Contributions}
D.B. and G.D. conceived the idea, designed the research, and supervised the work. 
N.R. and  A.P.R. performed simulations. All authors discussed the results and contributed to the final manuscript. D.B. and G.D. wrote the manuscript. 
 
 \section*{Competing Interests}
 The authors declare no competing interests.

\newpage


%
\pagebreak 

\end{document}